\documentclass[aps,prb,twocolumn,superscriptaddress,showpacs]{revtex4-2}
\usepackage{graphicx}  % Include figure files
\usepackage{amsmath}   % For advanced math
\usepackage{amssymb}   % For additional symbols
\usepackage{bm}        % For bold math symbols
\usepackage[colorlinks=true, linkcolor=blue, citecolor=blue, filecolor=blue, urlcolor=blue]{hyperref}
\usepackage{multirow} % Add this line to include the multirow package
\usepackage{textcomp} % In the preamble
\usepackage{subcaption}
\usepackage{caption}
\usepackage{array}
\usepackage{titlesec} % Package for customizing titles
\usepackage{comment}

\begin{document}

\newcommand{\Ef}{E\textsubscript{F}}

\title{Electronic and magnetic ground states of \{112\} grain boundary in graphene in the extended Hubbard model}

\author{Sishir Jana}
\affiliation{UGC-DAE Consortium for Scientific Research, DAVV Campus, Khandwa Road, Indore - 452001.}
\email{Corresponding author: rajamani@csr.res.in}
\author{Dayasindhu Dey}
\affiliation{Department of Physics, School of Advanced Sciences, Vellore Institute of Technology (VIT), Chennai Campus, Chennai 600 127 Tamil Nadu, India}
\author{Manoranjan Kumar}
\affiliation{S.N. Bose National Centre for Basic Sciences, Block-JD, Sector-III, Salt Lake, Kolkata - 700106.}
\author{S. Ramasesha}
\affiliation{Solid State and Structural Chemistry Unit, Indian Institute of Science, Bangalore - 560012.}
%\author{Ram Janay Choudhary}
%\affiliation{UGC-DAE Consortium for Scientific Research, DAVV Campus, Khandwa Road, Indore - 452001.}
\author{Rajamani Raghunathan}
\affiliation{UGC-DAE Consortium for Scientific Research, DAVV Campus, Khandwa Road, Indore - 452001.}

\date{\today}

\begin{abstract}
We study the ground state phase diagram of the extended
Hubbard model in a half-filled 5/7 skewed ladder, which is topologically 
equivalent to a \{112\} grain boundary in graphene and related systems.
Using the mean-field method, we identify various electronic and magnetic
phases in the U-V plane, by calculating the site charge and spin densities. 
The electronic phases include partially charge-ordered metal or insulator, 
and fully charge-ordered insulator. The different magnetic phases of the 
model are non-magnet, spin density wave, spin split compensated ferrimagnet or
partial antiferromagnet. Analysis of the 
electronic band structure reveals that the partially charge-ordered compensated
ferrimagnetic phase exhibits spin polarisation, which can be 
quite interesting for spintronics applications. We also compute 
the polarisation as a function of $U$ using the Berry phase 
formalism and show that the system 
exhibits multiferroicity with coexisting compensated ferrimagnetic spin order alongside 
electronic polarisations. 

\end{abstract}

\maketitle

\section{\label{sec:Intro}Introduction}
Modeling quantum phase transition (QPT) in low-dimensional systems 
is fundamentally significant, as it reveals novel electronic \cite{iemini2015}
and magnetic \cite{kumar2015} ground states within the model parameter space. 
The quantum fluctuations in low-dimensional systems are confined and 
give rise to many interesting quantum phases in the ground state \cite{ogino2022}. These fluctuations 
further determine the low temperature electronic  \cite{paki2019,pudleiner2019} and 
magnetic \cite{Sachdev1999} behaviours of the system. 
Many of these quantum systems exhibit charge \cite{chisa2006}  or spin frustration  \cite{Lacroix2011}
due to competing interactions, leading to several exotic quantum phases like the 
charge density wave(CDW) \cite{sarker2023}, spin density wave (SDW)\cite{mazumdar1999},
bond order wave (BOW) \cite{kumar2009}, spiral \cite{White1996} or dimer order \cite{Majumdar1969} etc. 
To model these quantum phases, extended Hubbard model  \cite{qu2022} and various types of Heisenberg models have been employed 
on one-dimensional chains \cite{bera2022}, triangular ~\cite{pandey2017} and square ladders \cite{zhou2023},
as well as skewed ladder systems  \cite{das12021,dey2020}.

\begin{figure}[htbp]
    \centering
    \begin{minipage}{0.90\linewidth}
        \centering
        \includegraphics[width=\linewidth]{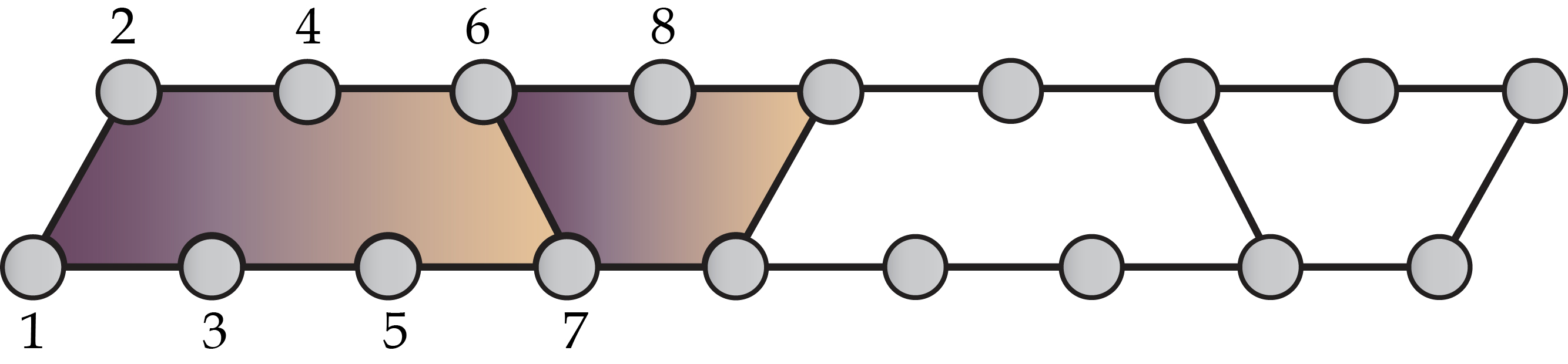}
    \end{minipage}%
\\\vspace{1em}
    \begin{minipage}{0.5\linewidth}
        \centering
        \includegraphics[width=\linewidth]{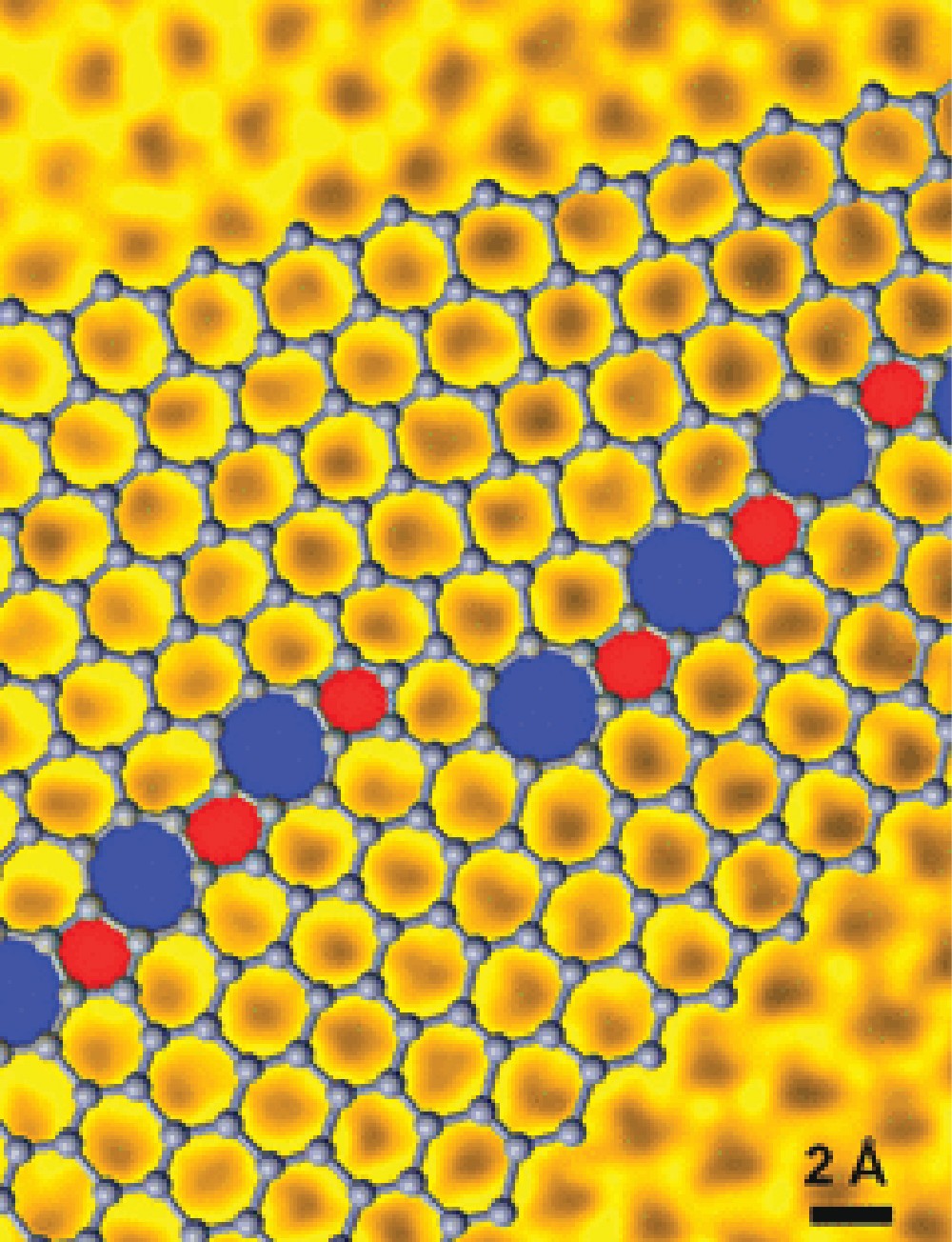}
    \end{minipage}
    \caption{\label{fig:unitcell} 
    (Top) Unit cell of a 5-7 skewed ladder system. 
    (Bottom) Scanning transmission electron microscopy (STEM) image of \{112\} grain boundary reproduced with permission from ~\cite{kim}. 
     Copyright 2011 ACS. The black spheres in the figure represent the carbon atoms. }
\end{figure}

\vspace{1em}
\noindent Skewed ladder systems consist of two-leg ladders with slanted rung interactions between the legs. 
Among these, the 5/7 skewed ladder which is shown in the figure \ref{fig:unitcell} is particularly 
interesting due to its resemblance to the structure of fused Azulenes and is also formed at the \{112\} grain boundary of graphene.
Using the Pariser-Parr-Pople (PPP) model, Thomas et al. 
demonstrated that fused Azulene can function as a room-temperature 
organic multiferroic material  ~\cite{Simil}. Giri and coworkers have also 
applied the PPP model to investigate singlet fission 
in polycyclic aromatic hydrocarbons  ~\cite{giri2019correlated}. 
More recently, Valentim et al. 
explored the magnetic phase diagram of the Hubbard model 
in a 5/7 ladder, revealing a system size (n)-dependent 
magnetic phase transition from a singlet (S = 0) to 
higher spin ground states (S = 1, 2) in the $U$ {\it vs}. $n$ phase space, 
which was first observed in the PPP model and the Heisenberg spin model by Thomas et al.  ~\cite{Garcia}.

%\begin{figure}[htbp]
    %\centering
    %\begin{minipage}{0.65\linewidth}
        %\centering
        %\includegraphics[width=\linewidth]{5-7UnitCell.jpg}
    %\end{minipage}%
    %\hspace{1em}
    %\begin{minipage}{0.3\linewidth}
        %\centering
        %\includegraphics[width=\linewidth]{stm1.jpg}
    %\end{minipage}
    %\caption{\label{fig:unitcell} 
    %(Left) Unit cell of a 5-7 skewed ladder system. 
    %(Right) Scanning transmission electron microscopy (STEM) image of \{112\} grain boundary from reference~\cite{kim}.}
%\end{figure}

\vspace{1em}
\noindent In addition to the electronic models, 
several studies have explored the magnetic phases in the ground 
state of skewed ladders using Heisenberg spin-1/2 and spin-1 models.
~\cite{giri2017,das12021}. Notably, 
systems such as 3/4, 3/5 and 5/7 skewed ladders exhibit 
a quantum phase transition (QPT) from non-magnetic to ferrimagnetic ground state. 
These QPT boundaries can be identified by analyzing the discontinuities 
in the entanglement entropy and sharp dips in the fidelity 
at the phase boundaries  ~\cite{das2022,das2024}. 
Additionally, the spin-1/2 5/7 skewed ladder is also known 
to exhibit unusual plateaus at 1/4, 1/2 and 3/4 magnetizations ~\cite{dey2020}.
Studies of the spin-1 5/7 skewed ladders have shown the presence of a vector 
chiral phase induced by the breaking of both reflection and parity symmetries in the ground state ~\cite{das12021,das22021}. 

\vspace{1em}
\noindent Beyond model systems, skewed ladder structures are also interesting 
due to their similarity to grain boundaries (GBs) commonly found in two-dimensional 
materials like graphene and bulk materials such as silicon.
%Recently, N.Sarker et. al. have studied the GB of graphene using 
%Scanning Tunneling Microscope measurement ~\cite{Sarker2025}.
Hsieh and coworkers, through 1/f noise studies on graphene GBs have 
demonstrated spontaneous time-reversal symmetry breaking, leading to
ferromagnetism ~\cite{hsieh2021}.
Investigations of various prevalent GBs in silicon have further revealed 
distinct electronic structures. Notably, the \{112\} GB, which 
closely resembles the 5/7 skewed ladder, exhibits a metallic 
ground state, whereas other GBs display a gaped nature ~\cite{rajamanisir2014}. These 
prior studies suggest that 5/7 skewed ladders could host a wide 
range of electronic and magnetic phases.

\vspace{1em}
\noindent In order to understand the electronic properties of GBs, 
it is essential to study 5/7 skewed ladder using models like 
the extended Hubbard model. The presence of both 
on-site ($U$) and inter-site ($V$) Coulomb repulsions 
can host more exotic properties due to the competition 
between the charge and spin degrees of freedom. 
We therefore investigate here the emerging 
electronic and magnetic phases on a 5/7 skewed ladder 
system using an extended Hubbard model at half-filling.

\vspace{1em}
\noindent In this work we analyze the extended Hubbard model 
using a mean-field approach. We construct the 
quantum phase diagram in the $U$ {\it vs}. $V$ parameter space, and also 
show that the phase diagram exhibits rich electronic and magnetic phases 
in the model parameter space including 
partially charge-ordered metal or
insulator, fully charge-ordered insulator, 
spin density wave metal and spin split compensated ferrimagnet (SS CFiM). 
These phases are characterized using the site spin and charge densities as 
well as the electronic band structures. Finally, we present the results 
of our polarization calculations and conclude. 

\section{\label{sec:Method}Model and Method}

\vspace{1em}
\noindent The extended Hubbard model that describes a 5/7 skewed ladder system is given by, 

%=====================================================================
%=====================================================================
\newcommand{\cdag}[1]{\hat{c}^{\dagger}_{#1}}
\newcommand{\can}[1]{\hat{c}_{#1}^{\phantom{\dagger}}}
\begin{eqnarray}
\label{eq:model}
\hat{\mathcal{H}}_{5/7} &=& -t \sum_{i=0}^{n-1} \sum_{j=1}^{8} \sum_{\sigma}
\left( \cdag{8i+j,\sigma} \can{8i+j+2,\sigma} + \text{h.c.} \right) \nonumber \\
&& - t \sum_{i=0}^{n-1} \sum_{\sigma}
\left( \cdag{8i+1,\sigma} \can{8i+2,\sigma} +
       \cdag{8i+6,\sigma} \can{8i+7,\sigma} + \text{h.c.} \right) \nonumber \\
&& + U \sum_{i=0}^{n-1} \sum_{j=1}^{8} \hat{n}_{8i+j,\uparrow} \hat{n}_{8i+j,\downarrow} \nonumber \\
&& + V \sum_{i=0}^{n-1} \sum_{j=1}^{8} \hat{n}_{8i+j} \hat{n}_{8i+j+2} + V \sum_{i=0}^{n} \hat{n}_{8i+1} \hat{n}_{8i+2}
\end{eqnarray}

\noindent where, $i$ labels the unit cell, $j$ represents the site within the unit cell and 
$n$ represents the number of unit-cells. 
The operators \( \hat{c}_{i,\sigma}^{\dagger} \) and \( \hat{c}_{i,\sigma} \) 
are the creation and annihilation operators at site $i$ with spin $\sigma$ respectively,  
$t$ is the hopping parameter, representing the kinetic 	
energy of electrons hopping between neighboring sites. The parameter 
\( U \) is the on-site Coulomb repulsion term, accounting for the energy 
cost of two electrons occupying the same site, 
\( V \) is the inter-site Coulomb repulsion term, representing the repulsion energy 
for electrons on neighboring sites. The operator 
\( \hat{n}_{i,\sigma} = \hat{c}_{i,\sigma}^{\dagger} \hat{c}_{i,\sigma} \) is 
the number operator for electrons with spin \( \sigma \) at site \( i \).
The total electron number operator at site \( i \) is given by 
\( \hat{n}_{i} = \sum_{\sigma} \hat{n}_{i,\sigma} \). 
In equation \ref{eq:model} the terms in the first two lines correspond to the 
leg and rung hopping, the third line accounts for the on-site Coulomb repulsion at site $j$ 
and the inter-site Coulomb repulsions forms the last line. 
%=====================================================================
%\begin{eqnarray}
   %\label{eq:model}
    %\hat{H} & = & -\sum_{\langle i,j \rangle, \sigma} t_{ij} 
    %\left(\hat{c}_{i,\sigma}^{\dagger} \hat{c}_{j,\sigma} + 
     %\hat{c}_{j,\sigma}^{\dagger} \hat{c}_{i,\sigma}\right) \nonumber \\
    %&& + U \sum_{i} \hat{n}_{i,\uparrow} \hat{n}_{i,\downarrow} 
    %+ V \sum_{\langle i,j \rangle} \hat{n}_{i} \hat{n}_{j} ,
%\end{eqnarray}

\vspace{1em}
\noindent The real-space Hamiltonian in the equation \ref{eq:model} can be 
transformed into the reciprocal or $\mathbf{k}$ space ($\hat{\mathcal{H}}_\mathbf{k}$) by doing 
a discrete Fourier transform of the creation and annihilation operators as shown in the equation \ref{eq:oprtransform}.

\begin{eqnarray}
\label{eq:oprtransform}
	\hat{c}^{\dagger}_{i,\sigma} &= & \frac{1}{\sqrt{N}} \sum_{\mathbf{k}} e^{-i \mathbf{k} \cdot \mathbf{r}_i} \, \hat{c}^{\dagger}_{\mathbf{k},\sigma} \nonumber \\
	\hat{c}_{i,\sigma} &= & \frac{1}{\sqrt{N}} \sum_{\mathbf{k}} e^{i \mathbf{k} \cdot \mathbf{r}_i} \, \hat{c}_{\mathbf{k},\sigma} 
\end{eqnarray}

\vspace{1em}
\noindent where $N$ is the number of atoms in the unit-cell. The $\mathbf{k}$ space Hamiltonian, $\hat{\mathcal{H}}_\mathbf{k}$ obeys the 
relation shown in the equation \ref{eq:Hamiltonian_k}.
%========================
\begin{eqnarray}
   \label{eq:Hamiltonian_k}
\hat{\mathcal{H}} &=& \sum_{\mathbf{k}} \psi_{\mathbf{k}}^\dagger \hat{\mathcal{H}}_{\mathbf{k}} \psi_{\mathbf{k}} \nonumber \\
\hat{\mathcal{H}}_{\mathbf{k}} &=& \hat{\mathcal{H}}_{\mathbf{k}}^\uparrow \oplus \hat{\mathcal{H}}_{\mathbf{k}}^\downarrow \nonumber 
\end{eqnarray}

\begin{eqnarray}
\text{and} \quad \mathbf{\psi}_{\mathbf{k}} = 
\begin{pmatrix}
c_{1\mathbf{k}\uparrow} \\ c_{2\mathbf{k}\uparrow} \\ c_{3\mathbf{k}\uparrow} \\ c_{4\mathbf{k}\uparrow} \\
c_{5\mathbf{k}\uparrow} \\ c_{6\mathbf{k}\uparrow} \\ c_{7\mathbf{k}\uparrow} \\ c_{8\mathbf{k}\uparrow}
\end{pmatrix}
\otimes
\begin{pmatrix}
c_{1\mathbf{k}\downarrow} \\ c_{2\mathbf{k}\downarrow} \\ c_{3\mathbf{k}\downarrow} \\ c_{4\mathbf{k}\downarrow} \\
c_{5\mathbf{k}\downarrow} \\ c_{6\mathbf{k}\downarrow} \\ c_{7\mathbf{k}\downarrow} \\ c_{8\mathbf{k}\downarrow}
\end{pmatrix}
\end{eqnarray}
%=================================

\vspace{1em}
\noindent The model is solved using a mean-field approach \cite{Claveau} in which 
the operator $\hat{n}_i$ is written as the
sum of its expectation value \( \langle \hat{n}_i \rangle \) 
and its fluctuation \( \Delta n_i \), 
$\hat{n}_i \approx \langle \hat{n}_i \rangle + \Delta n_i$.

\begin{figure}
    \includegraphics[width=\columnwidth]{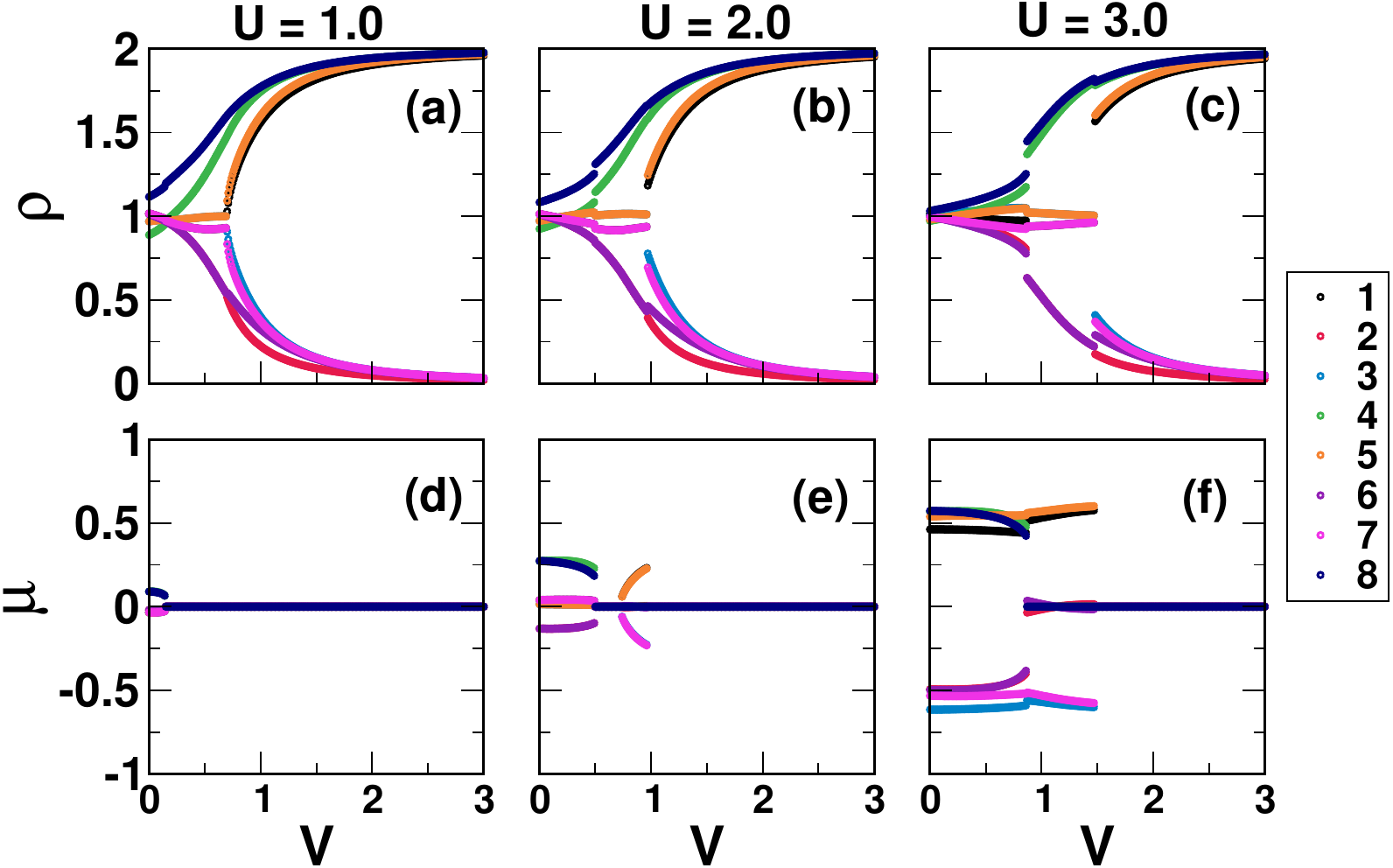}
    \caption{\label{fig:sp-chr-den} Variation of site charge (a-c) and spin (d-f) 
		densities as a function V for U=1.0, 2.0 and 3.0. There is a minor 
		charge difference in charge densities of sites 4 and 8 (upper-leg) 
		even at $t$=0. The CO transitions are continuous at lower-$U$ and 
		discontinuous in the large $U$ limit. The magnetic transitions are 
		by and large discontinuous.}
\end{figure}

\begin{table*}
    \begin{tabular}{|c|c|c|c|c|}
        \hline
        Operator & \(E\) & \(\sigma\) & $\mathcal{P}$ & $\sigma\mathcal{P}$ \\ \hline
        \(S_i^x\) & \(S_i^x\) & \(\begin{cases} S_{N-i}^x & \text{if } i \neq N, N/2 \\ S_i^x & \text{if } i = N, N/2 \end{cases}\) & \(-S_i^x\) & \(\begin{cases} -S_{N-i}^x & \text{if } i \neq N, N/2 \\ -S_i^x & \text{if } i = N, N/2 \end{cases}\) \\ \hline
        \(S_i^y\) & \(S_i^y\) & \(\begin{cases} S_{N-i}^y & \text{if } i \neq N, N/2 \\ S_i^y & \text{if } i = N, N/2 \end{cases}\) & \(S_i^y\) & \(\begin{cases} S_{N-i}^y & \text{if } i \neq N, N/2 \\ S_i^y & \text{if } i = N, N/2 \end{cases}\) \\ \hline
        \(S_i^z\) & \(S_i^z\) & \(\begin{cases} S_{N-i}^z & \text{if } i \neq N, N/2 \\ S_i^z & \text{if } i = N, N/2 \end{cases}\) & \(-S_i^z\) & \(\begin{cases} -S_{N-i}^z & \text{if } i \neq N, N/2 \\ -S_i^z & \text{if } i = N, N/2 \end{cases}\) \\ \hline
        \(c_{i,\uparrow}\) & \(c_{i,\uparrow}\) & \(\begin{cases} c_{N-i,\uparrow} & \text{if } i \neq N, N/2 \\ c_{i,\uparrow} & \text{if } i = N, N/2 \end{cases}\) & \(c_{i,\downarrow}\) & \(\begin{cases} c_{N-i,\downarrow} & \text{if } i \neq N, N/2 \\ c_{i,\downarrow} & \text{if } i = N, N/2 \end{cases}\) \\ \hline
        \(c_{i,\downarrow}\) & \(c_{i,\downarrow}\) & \(\begin{cases} c_{N-i,\downarrow} & \text{if } i \neq N, N/2 \\ c_{i,\downarrow} & \text{if } i = N, N/2 \end{cases}\) & \(-c_{i,\uparrow}\) & \(\begin{cases} -c_{N-i,\uparrow} & \text{if } i \neq N, N/2 \\ -c_{i,\uparrow} & \text{if } i = N, N/2 \end{cases}\) \\ \hline
    \end{tabular}
    \caption{Transformation properties of various spin and electronic operators under the symmetry operations, $E$, $\sigma$, $\mathcal{P}$ and $\sigma\mathcal{P}$.}
    \label{tab:symmetry}
\end{table*}

\noindent The system is initialised with suitable spin ($\mu$) and charge densities ($\rho$) 
on every site to represent a chosen electronic and magnetic state. 
The Hamiltonian matrix for the up- and the down-spin electrons is constructed in the 
$\mathbf{k}$ space ($\hat{\mathcal{H}}_\mathbf{k}^\sigma$) and the same is 
presented in the supporting information (SI) file.  
This Hamiltonian is diagonalised to obtain the eigenvalues and the corresponding 
eigenvectors. The $\mu$ and $\rho$ are then computed from the 
eigenvector, which form the inputs for the next iteration. 
The iterative process is repeated until a chosen convergence criterion 
on the charge density difference, $\delta\rho$ = 10$^{-5}$ is reached. 
In order to obtain the correct ground state, the system is also 
started with different initial electronic and magnetic configurations by altering 
the starting site charge and spin densities respectively. The energies of such configurations 
are compared to determine the electronic and magnetic ground states. 
The converged ground state wave functions are then used to 
calculate the band structure, density of states(DOS) as well as 
charge and spin densities as a function of \( U \) and \( V \) in order to characterize
different phases. The value of the hopping parameter ($t$) in 
the Hamiltonian in equation \ref{eq:model} is kept 
fixed at 1 eV, which serves as the energy scale for 
the parameters $U$ and $V$. The values of the model parameters $U$ and $V$ 
are varied in the range 0-6$t$ and 0-3$t$.

\section{\label{sec:RandD}Results}
\subsection{\label{ssec:sym} Symmetry of the 5/7 Skewed Ladder}
The unit cell of our 5/7 skewed ladder consists of eight sites, with
each site possessing on an average one electron in one orbital.
This corresponds to the half-filled skewed ladder. The sites are numbered
such that the lower leg sites are odd numbered, and the upper leg has 
even numbered sites as shown in figure \ref{fig:unitcell}. The connectivities 1-2 and 6-7
form the rung bonds. This unit cell can be repeated along the -$x$ and +$x$
directions to obtain the entire ladder. The unit cell of our model has
few inherent symmetries. These include (i) a mirror $\sigma$ with its plane
passing along sites 4 and 8, or a $C_2$ symmetry element whose rotation axis passes through 
sites 4 and 8 (ii) a parity or spin-inversion symmetry $\mathcal{P}$ which corresponds to
inversion of the spin from $+s$ to $-s$ or vice-versa
and (iii) a combination of the the first two symmetries $\sigma$ and $\mathcal{P}$.
Therefore, the symmetry elements of the ladder can
be written as $\{E, \sigma, \mathcal{P}, \sigma\mathcal{P}\}$, where $E$
is the identity operator. Action of these symmetry elements on different
fermionic and bosonic operators for $N$ = 8 sites is shown in the table \ref{tab:symmetry}.

\begin{figure*}[t]
    \includegraphics[width=\linewidth]{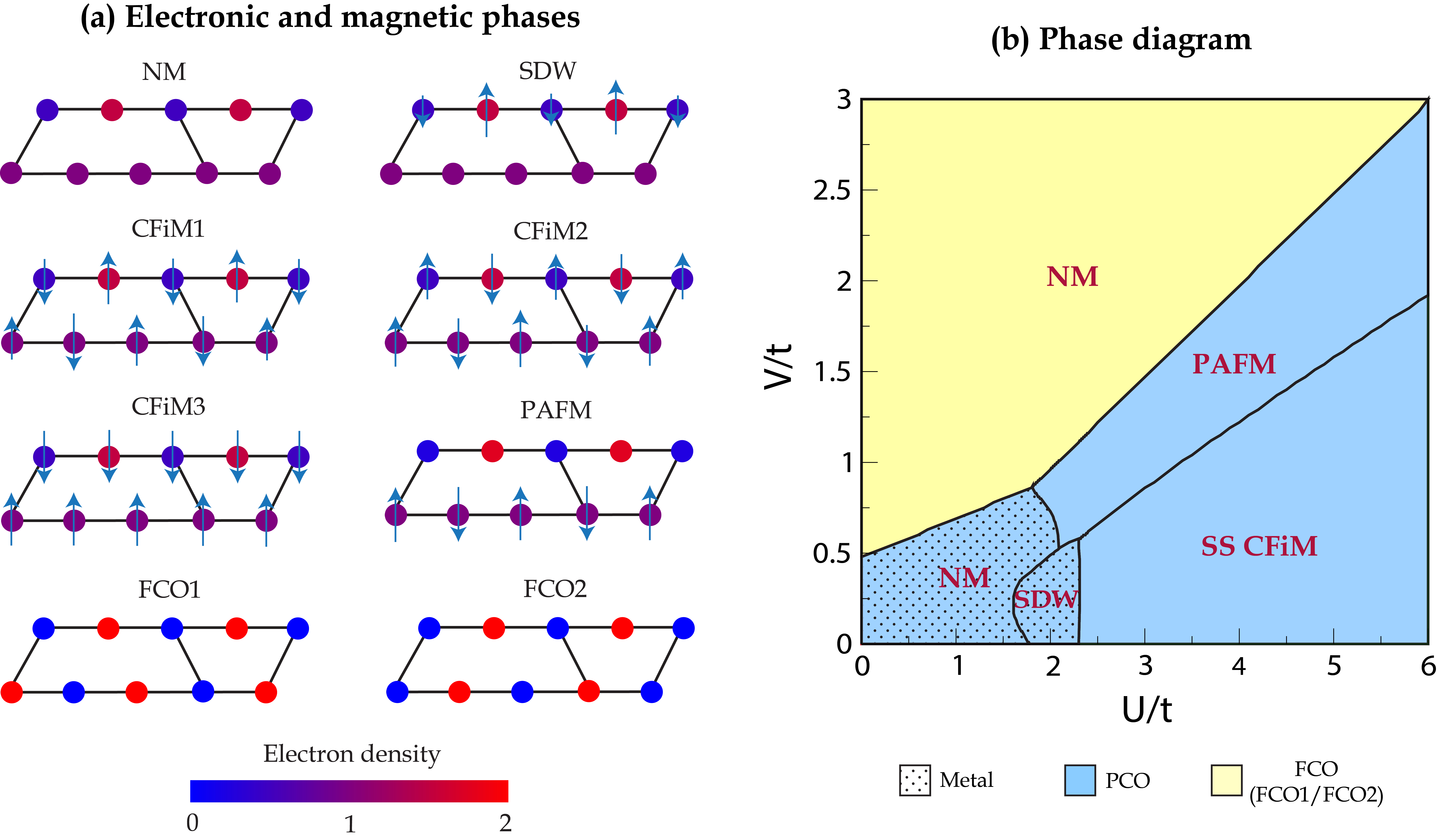}
	\caption{\label{fig:phasediagram} (a) Schematic of various electronic and magnetic phases of the model. 
		The color scale from blue through red represent the value of charge density $\rho$ on each site. 
                A perfect blue/red color corresponds to $\rho$ = 0 (empty) or 2 (doubly occupied) respectively. 
		The up- and down-spins are represented by the direction of the arrows and the length of the 
		arrow represents the magnitude. (b) Phase diagram of the 5/7 ladder 
		the $U$ {\it vs}. $V$ plane. The blue shaded regions are partially 
		charge-ordered phases and the yellow region corresponds to the fully CO (FCO) phase. The dotted regions 
		are the metallic phases and the un-dotted regions are gaped phases. The various phases seen in the 
		phase diagram are the PCO+NM (Blue+dot), PCO+SDW (Blue+dot), PCO+SS CFiM (Blue), PCO+PAFM (Blue) 
		and FCO+NM (yellow).}
\end{figure*}

\begin{figure*}[t]
    \includegraphics[width=\linewidth]{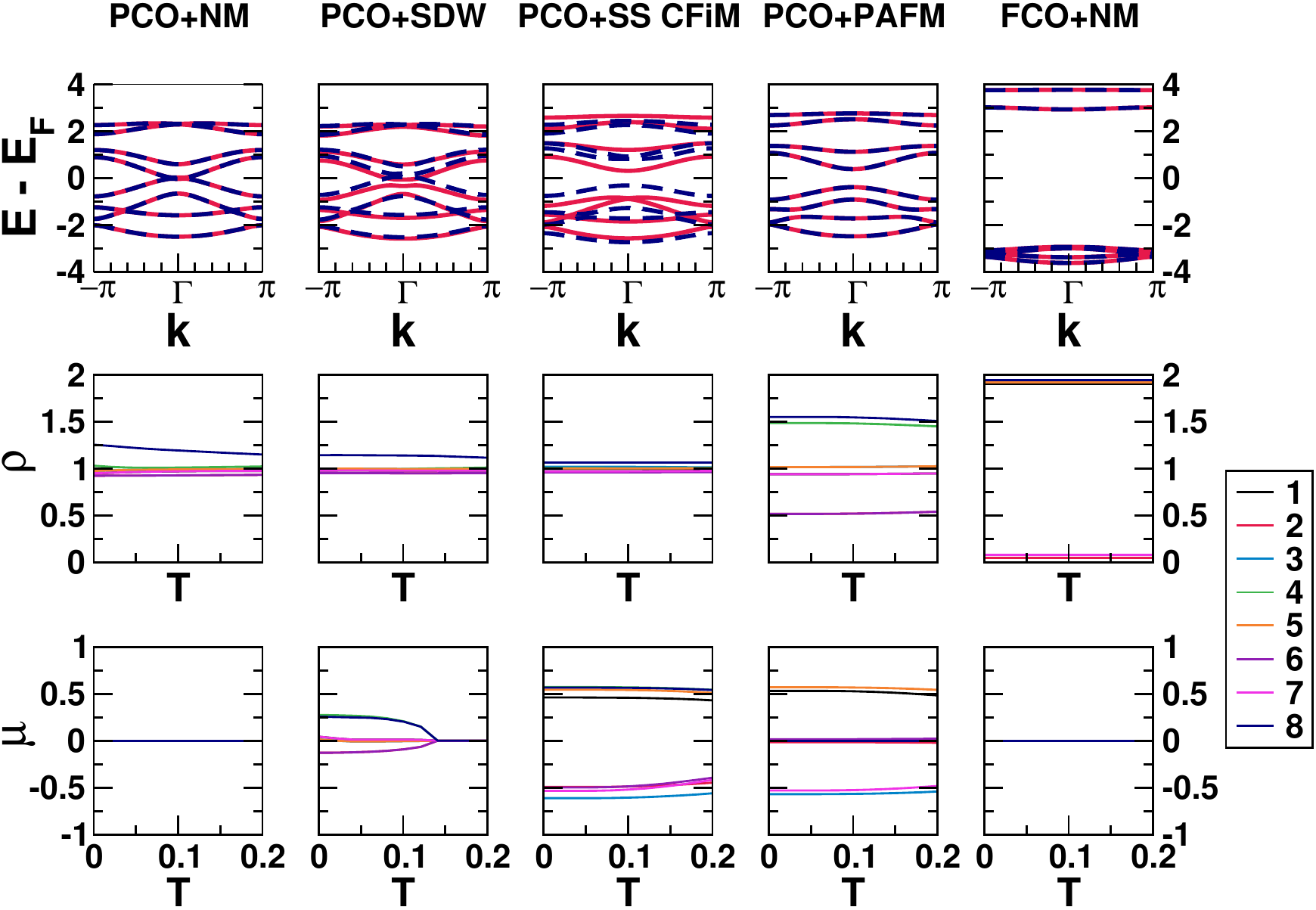}
	\caption{\label{fig:BS-Tdep} (Top row) Band structure in PCO+NM 
		PCO+SDW, PCO+SS CFiM, PCO+PAFM and FCO+NM phases of the model 
		from $k$ = -$\pi$ to +$\pi$. The Fermi level is fixed at 0 eV.
		The up- and down-spin channels are plotted separately in 
		red (continuous) and blue (dashed) lines. 
		(Middle row) Charge density, $\rho$ as a function of temperature in the 
		range 0-2$t$. (Bottom row) Temperature (in units of $t/k_B$) dependence of spin density $\mu$ 
		in different phases of the model.  }
\end{figure*}

%===================================================
\begin{figure*}[t]
    \includegraphics[width=\linewidth]{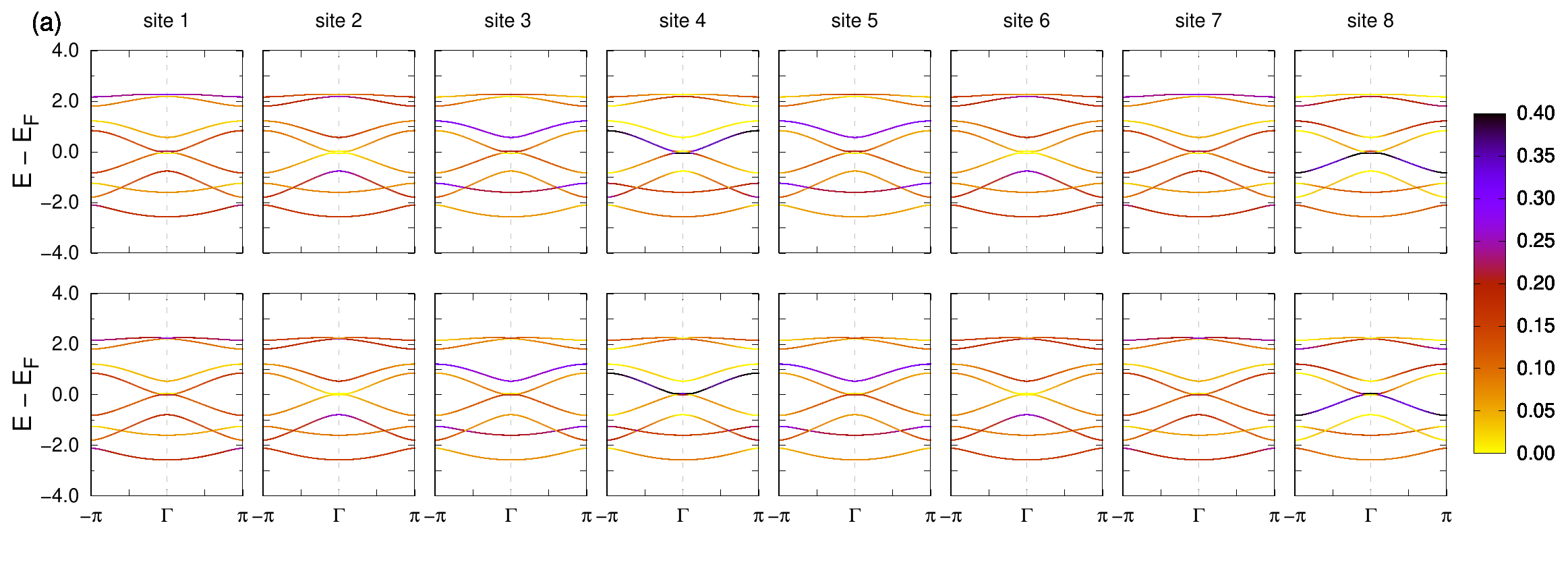}
    \includegraphics[width=\linewidth]{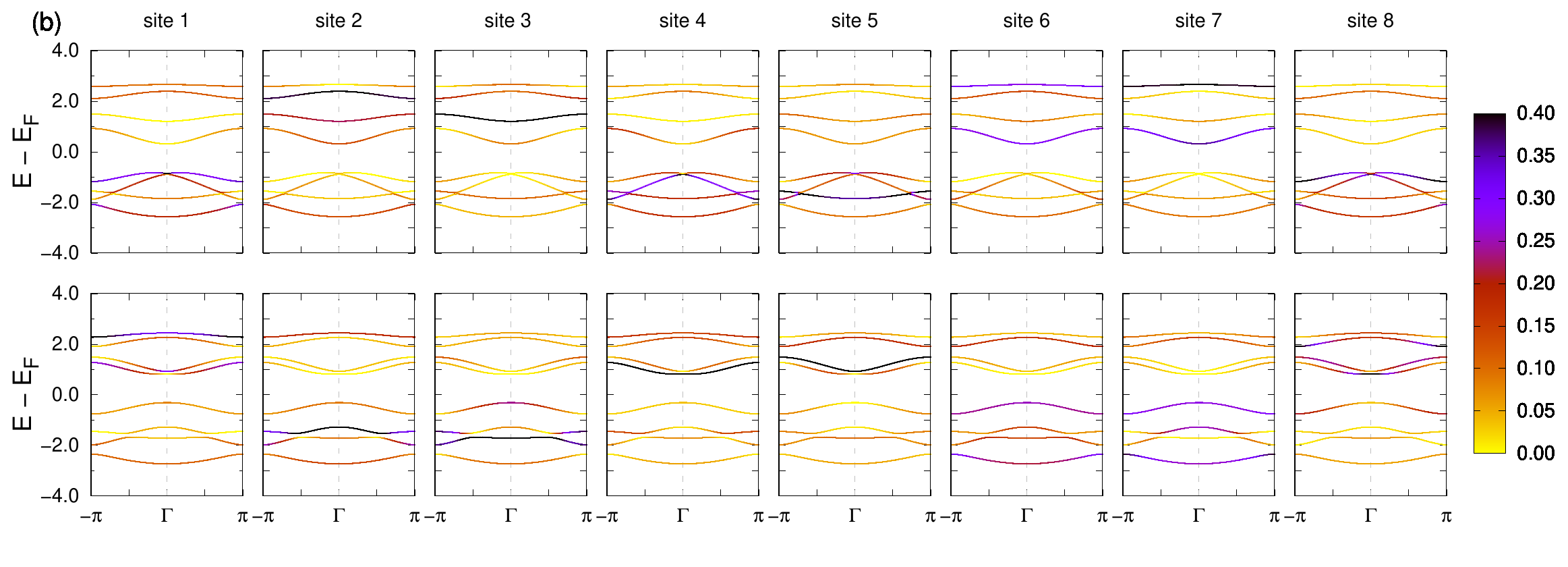}\\
	\caption{\label{fig:ProjBS} Site projected electronic band structure of the (a) PCO+NM and (b) PCO+SS CFiM
	phases from $k$ = -$\pi$ to +$\pi$. The Fermi level is fixed at 0 eV. 
	The top and bottom panels in (a) and (b) correspond to the up- and down-spin channels. 
	The color scale on the right represents the weight or contribution of the particular site for a 
	given band at different $k$-points. For example a band with black color represents maximum contribution 
	by the site. }
\end{figure*}
%===================================================
\subsection{\label{ssec:ChrDen} Charge density}

First we discuss the site charge density $\rho$ of the model for 
different V, which is shown in the figure \ref{fig:sp-chr-den} a-c. 
It would be instructive to understand the $t=0$ behaviour. 
A large value of $U$ will stabilise those states which have 
single occupancy of the orbitals. On the other hand the 
parameter $V$ adds an energy cost when the orbitals on the neighbouring sites are 
both occupied, thereby favouring double occupancy on the alternate sites. 
Therefore, the $U$ and $V$ parameters stabilise 
those states with singly or doubly occupied orbitals respectively. 
In this limit the charge ordering will occur whenever $V\ge U/2$. 
Now we turn our attention to the $t\ne 0$ behaviour.
The 5/7 ladder hosts mirror symmetry $\sigma$, the plane of which 
passes through sites 4 and 8, and perpendicular to the plane of the ladder. 
Hence, the charge densities of
the sites 1 and 7, 3 and 5 as well 2 and 6
are the same in those phases where $\sigma$ or $\sigma\mathcal{P}$ symmetry is conserved.  
For non-zero but very small hopping parameter, we notice that there is 
a subtle charge difference between the sites 4 and 8 even for $V<U/2$. 
Therefore we calculated $\rho$ in the 
$t\rightarrow 0$ limit ($t$=0.001), in which case there is no 
charge difference between these two sites for $V<U/2$. Also, when  
we turn off the hopping parameter for the rung bonds,   
keeping the $t$ of leg bonds non-zero, the site charge differences 
are negligible. Hence we conclude that the 
charge difference between the sites 4 and 8 
in the $t\ne 0$ limit originates from the 
rung bonds. We further notice that the charge 
difference is large when the value of $U$ is less. The charge densities of 
different sites in the upper and lower legs begin to bifurcate 
at higher values of $V$ leading to charge-order (CO). Moreover, the 
onset of bifurcation in the upper leg occurs at a lower value of 
$V$ compared to the lower leg for a $U$ value. For any given 
$U$ value, when the inter-site repulsion, $V$ is 
weak, the upper leg shows charge-ordering while the lower leg 
shows uniform charge density. 
We refer to this as the partially charge-ordered (PCO) phase. For instance, 
when $U$ = 1 and  $V$ = 0.25, $\rho$ of sites 4 and 8 start increasing with $V$ 
whereas the charge density of sites 2 and 6 decrease, however 
the charge density of sites in the lower leg remain uniform and close 
to 1. 
A similar trend is observed for higher $U$ values also. This could be 
due to the difference in the topology of the upper and lower leg. 
The upper leg has three sites between two rungs while the lower leg has 
either two or four sites between any two rungs. The odd number of sites
between rung bonds leads to frustration in the local rings and hence
delocalisation is reduced therefore the system prefers a charge ordered state.
For large values of $V$ we notice that both the upper and lower legs
are charge-ordered, which we refer as the fully charge-ordered (FCO) phase. 
The FCO phase is doubly degenerate with charge configurations FCO1 and FCO2, 
due to the presence of mirror symmetry.

\subsection{\label{ssec:SpnDen} Magnetic moment}
The magnetic moment of site $i$ ($\mu_i=(n_{i\uparrow}-n_{i\downarrow}$)) computed as a function of $V$ 
at different $U$ values is presented in figure \ref{fig:sp-chr-den} d-f. 
The plot reveals that for small $U$, the magnetic moment of all the eight sites are negligibly small, 
corresponding to a non-magnetic (NM) state. When $U$ is increased to 2.0, the sites 
on the upper leg show non-zero moments while the sites on the lower leg 
have nearly zero moment. This can again be attributed to the magnetic frustration 
in the upper leg. In the lower leg the spin on initial site almost determines 
the spin on other sites. Furthermore, the spin at the initial site of
the leg can be up or down, leading to zero mean spin density. We further notice
that at $V$ = 0, the spin magnetic moments of sites 4 and 8 are 
large positive (about +0.25 $\mu_B$) and those of sites 2 and 6 are relatively 
smaller and also negative (about -0.15 $\mu_B$). This means that the upper leg shows SDW,
while the lower leg is non-magnetic. When $V$ is increased gradually, 
the SDW order of the upper leg ceases and the entire ladder develops a non-magnetic (NM) phase.
Upon further increasing $V$ an antiferromagnetic order develops in the lower leg, while the 
upper leg still remains non-magnetic, which we call the partially antiferromagnetic (PAFM) phase. 
Finally the antiferromagnetic order in the lower leg also melts into a NM phase at high $V$ values. 

\vspace{1em}
\noindent For $U>3$ and $0 \le V \le 0.9$, we 
observe anti-parallel alignment of magnetic moments 
in both the legs, with moment value close to 0.5 $\mu_B$. The calculated 
spin densities in this region reveal that the site moments 
are opposite in direction but not equal in magnitude, like the 
ferrimagnets. However, the total magnetic moment of the 
unit-cell is zero similar to the conventional antiferromagnets. 
This type of magnetic ordering is unique in the sense that 
this magnetic phase has the characteristics of both ferrimagnets and 
antiferromagnets. We will later show that the electronic structure of this 
phase shows spin polarisation of the up- and down-spin bands. Therefore, 
we designate this phase as the partially charge-ordered, spin split (SS) 
compensated ferrimagnet (CFiM),  PCO + SS CFiM. 
The magnetic ground state of this phase consists of two unique 
magnetic configurations, CFiM1 and CFiM2 as shown in the figure \ref{fig:phasediagram}. Each of these magnetic 
configurations are doubly degenerate, thus making the ground state to 
be four-fold degenerate. Moreover, the CFiM1 and CFiM2 configurations are related by 
a combination of $\sigma$ and $\mathcal{P}$ symmetry. 
The other possible spin configuration, CFiM3 
is higher in energy. This can be 
easily understood from the energies of the singlet and triplet pairs in the localisation 
limit, which are respectively -3/4$J$ and 1/4 $J$. For the 
CFiM1 and CFiM2  phases, out of the total eight bonds, seven are antiferromagnetically coupled 
while one bond is ferromagnetic, yielding a total energy of -5$J$. 
However, in the CFiM3 phase two bonds are antiferromagnetic and six bonds  
are ferromagnetic, with total energy 0$J$. Hence, the CFiM1 and CFiM2 phases are 
degenerate and lower in energy than the CFiM3 phase.

\vspace{1em}
\noindent Increasing $V$ at large $U$ leads to PCO + SS CFiM to PCO + PAFM phase transition in 
which the upper leg becomes non-magnetic due to the onset of charge-order in that leg, 
while the lower leg retains the antiferromagentic order. Upon further 
increasing $V$ the system transforms to a NM phase due to the 
charge-ordering in both the legs. 
We also observe that for a given $V$, the magnetic moment increases 
with increasing $U$ value due to the enhanced localisation. The 
results of the charge and spin densities of the model can be 
combined to identify different phases of the system in the 
model parameter space of $U$ and $V$ namely PCO+NM, PCO+SDW, PCO+SS CFiM, PCO+PAFM 
and FCO+NM, as depicted in the figure \ref{fig:phasediagram}.

\subsection{\label{ssec:ElectronicStructure}Electronic Structure}
We now discuss the electronic band structures of the different phases of the model as 
presented in figure \ref{fig:BS-Tdep}. The bands corresponding to $\uparrow$ and 
$\downarrow$ spin are plotted as red-continuous and blue-dashed lines respectively. 
The electronic band structures of different phases are obtained in the middle of the 
phase, far enough from the phase boundary. The unit cell of the 5/7 skewed ladder consists of 
eight sites and the 
electronic band structure exhibits eight bands each in up- and down-spin channels. 
We observe that the phases corresponding to low $U$ and $V$ values namely PCO+NM and PCO+SDW 
are metallic in nature, depicted as dotted region in the phase diagram of 
figure \ref{fig:phasediagram}(c). The electronic bands of these two phases 
are very similar, except for the spin polarization between the 
bands corresponding to the up- and the down-spin channels in the SDW phase. 
The FCO+NM, PCO+PAFM and PCO+SS CFiM phases show a 
gaped behaviour. The electronic band gap of the FCO phase is the 
largest. The FCO+NM and PCO+PAFM phases do not show any spin polarization, 
but the PCO+SS CFiM phase shows considerable spin polarization. The presence of 
spin polarization in PCO+SS CFiM phase is quite interesting as the electronic 
bands of antiferromagnets are generally spin degenerate. However, 
momentum dependent spin polarization has been reported recently in 
certain unconventional antiferromagnets known as the altermagnets ~\cite{krempasky2024}. 
This is because, in altermagnets the neighbouring magnetic ions are related by spin-inversion plus an 
additional real space rotation symmetry ~\cite{smejkal}. On the contrary, in 
conventional antiferromagnets, 
the neighbouring sites are related by spin-inversion plus a real space translation 
symmetry ~\cite{smejkal}. For this reason, conventional AFMs do not show 
spin polarisation. But in case of our 5/7 skewed ladder, the spin 
polarization results from the unequal and opposite spin densities on different sites, 
while the total spin density of all the sites of the unit cell is zero. 
This magnetic ordering is different from ferrimagnetic ordering 
in the sense that even though the magnetic moments on different sites are unequal, the 
total magnetic moment of the unit-cell is zero. 
Such spin polarization in antiferromagnetic materials can be interesting for 
spintronics applications as they can exhibit spin Hall effect~\cite{jungwirth,nagaosa,ifmmode} . 

\vspace{1em}
\noindent In order to gain insights into the origin of these electronic bands, we 
have calculated the site projected band structure for each lattice site and the 
same is presented in the figure \ref{fig:ProjBS}. 
However, to illustrate the difference in the electronic structure of the 
metallic and gaped phases, we have presented the projected bands of only the PCO+NM and 
PCO+SS CFiM phases. The projected bands of the remaining three phases namely, 
FCO+NM, PCO+PAFM and PCO+SDW have been shown in the supporting information file. 
The contributions of different sites in the up- and down-spin channels are 
shown separately in the top and bottom rows respectively. 
First, we will discuss the projected band structure of the 
PCO+NM phase which is shown in figure \ref{fig:ProjBS} (a). 
In this phase the $\sigma$, $\mathcal{P}$ and $\sigma\mathcal{P}$ symmetries 
are preserved in the ground state. This results in identical band structures for sites 1 - 7, 2 - 6, 
and 3 - 5 for both the spin channels. 
Here, the band immediately below the Fermi level (\Ef) in the 
up- and down-spin channels is dominated by contributions from site 8. 
The band above \Ef~is contributed predominantly by site 4. We also 
notice band inversion happening at the $\Gamma$ point, at which the 
weights of the site 4 is transferred to bands below the \Ef. A similar observation 
is made in case of site 8. Furthermore, since this phase is non-magnetic, the 
up- and down-spin bands and their contributions of each and every sites are identical, 
due to the conservation of $\mathcal{P}$ symmetry. 
We also notice that there is a spin inversion plus band inversion symmetry between 
sites 4 and 8. For instance, the up-spin band above \Ef~ for site 4 is similar to the 
down-spin band below \Ef~for site 8 and vice-versa.

\noindent We now discuss the band structure of the PCO+SS CFiM phase. 
It should be recalled that the ground state of PCO+CFiM phase is degenerate, with 
CFiM1 and CFiM2 magnetic configurations. The CFiM1 and CFiM2 
magnetic states are related by $\sigma\mathcal{P}$ symmetry operation. 
Hence, in figure \ref{fig:ProjBS} (b), we only present the projected band structure of the 
CFiM1 phase. In this phase,  
bands of the up- and down-spin channels are not identical, giving rise to a 
net spin polarization. The up-spin occupied bands are 
mainly contributed by sites 1, 4, 5 and 8, 
while the sites 2, 3, 6 and 7 contribute via the down-spin 
channel, due to the CFiM1 type AFM interaction. 
It should be noted that the CFiM1 phase does not possess 
any symmetry and hence the contributions from different sites in 
up- and down-spin channels have no correspondence.

\subsection{Temperature dependence}
We have also studied the temperature dependence of charge and spin
density of the five different phases in the range $0\le$T$\le 0.2$ 
and the results are presented in 
figure \ref{fig:BS-Tdep}. The temperatures are in units of $t/k_B$, where 
$k_B$ is the Boltzmann's constant. 
Both the charge and spin densities 
undergo a continuous or second-order phase transition as a function of 
temperature. We observe that the charge and spin orderings of the PCO+PAFM and FCO+NM phases 
which occur at high value of $U$ or $V$ are quite robust, 
and we do not observe any transition in the 
temperature range that we have studied. 
The PCO+NM phase exhibits a gradual melting of the charge-order. 
The remaining two phases namely, the PCO+SDW and 
PCO+SS CFiM show a subtle charge-order in the entire temperature range 
that we have studied. The SDW phase becomes NM above T = 0.14. 
The SS CFiM phase also does not show any transition and hence 
is stable even at high temperatures. This is crucial for room temperature 
spintronic applications.

%===================================================
\subsection{Berry Phase}
Berry phase is the quantity that describes how the global phase 
accumulates when the complex wave function
is carried around a closed loop, in the parameter space in general. 
In our case, the 
closed loop is carried out in the reciprocal or $\mathbf{k}$ space.
Lately, the Berry phase has widely been calculated ~\cite{Daggato} 
on closed two-dimensional manifold, to obtain 
Chern number, which is an index for the topological nature 
of the surface S. In case of the one-dimensional systems 
the berry phase is also known as the Zak phase ~\cite{zak} which 
is given by, 

\begin{eqnarray}
\label{eq:BP-resta}
\gamma = i \sum_{n=1}^{n_b} \int_{C} \langle u_{n\mathbf{k}} | \nabla_{\mathbf{k}} u_{n\mathbf{k}} \rangle \cdot d\mathbf{k}
\end{eqnarray}

\noindent In equation \ref{eq:BP-resta}, \( |u_{nk}\rangle \) is the 
cell periodic part of the Bloch function, and the integration
is carried over a closed loop. Calculating the berry phase 
using the above expression is fairly straightforward, provided 
the phase of the Block function is smooth with respect to $\mathbf{k}$ .
However, in our case, since we diagonalise $\hat{\mathcal{H}}_\mathbf{k}$ 
at discrete values of $\mathbf{k}$, the wave functions at $\mathbf{k}$ 
and $\mathbf{k}+\Delta\mathbf{k}$ is not smooth, and differ by a random phase factor. 
Additionally, the wave function is not well defined wherever there 
is band degeneracy. Therefore, the calculation of the berry phase 
using the equation  \ref{eq:BP-resta} can give spurious results. 

\vspace{1em}
\noindent Alternate ways to calculate berry phase for the single band systems 
have also been proposed in the literature. There, random phase between 
the wave functions at $\mathbf{k}$ and $\mathbf{k}+\Delta\mathbf{k}$ 
cancels out when taking the product of the bra and ket states. In this case, 
the Berry phase is given by,~\cite{resta2000,resta2020}

\begin{eqnarray}
\label{eq:BP-SB}
 \gamma &=& - \sum_n \text{Im ln} \left[ \langle u_{n\mathbf{k}_0} | u_{n\mathbf{k}_{1}} \rangle \langle u_{n\mathbf{k}_1} | u_{n\mathbf{k}_{2}} \rangle \langle u_{n\mathbf{k}_2} | u_{n\mathbf{k}_{3}} \rangle  \right. \nonumber\\
&& \left. \cdots \langle u_{n\mathbf{k}_{N-1}} | e^{-\frac{2\pi ix}{a}}|u_{n\mathbf{k}_{0}} \rangle   \right].
\end{eqnarray}

\noindent In case of bands with degeneracy at certain $\mathbf{k}$-points, we need to use a multi-band 
approach, in which case the berry phase is calculated using the the relation~\cite{vanderbilt} shown below. 

\begin{eqnarray}
\label{eq:BP-MB}
&&\gamma = -\operatorname{Im} \log \left( \prod_{s=0}^{M-1} \det S(\mathbf{k}_s, \mathbf{k}_{s+1}) \right) \nonumber\\
&&S_{nn'}(\mathbf{k}_s, \mathbf{k}_{s+1}) = \langle u_{n\mathbf{k}_s} \,|\, u_{n'\mathbf{k}_{s+1}} \rangle
\end{eqnarray}

\vspace{1em}
\noindent where, $S_{nn'}(\mathbf{k}_s, \mathbf{k}_{s+1})$ is the overlap matrix. 
The polarization of the system in terms of the Berry phase is given by,  

\begin{equation}
P = \frac{\gamma}{2\pi} \, (\text{mod } e).
\end{equation}

%\noindent In systems with inversion symmetry, the Berry phase 
%is quantized to either \( 0 \) or \( \pi \), 
%corresponding to trivial or non-trivial topological phases, respectively.
%In the nontrivial phase there is emergence of topologically protected edge 
%states ~\cite{Chen2014}. We at first calculate Berry phase with only tight binding Hamiltonian and
%we get non zero value of berry phase. Berry phase for lowest four up and down band in unit of pi 
%is following.
   %up band  :  0.00000   0.25000   0.50000   0.75000
  %down band :  0.00000   0.25000   0.50000   0.75000
%
%Berry phase is not quantized in unit of $\pi$ which due to absence of 
%symmetry protected topological phase ~\cite{xiao2016,shang2024}. Another way to verify if it is topological or not 
%we calculate single particle energy spectrum in non-interacting tight binding case and we do not find
%any topologically protected edge state ~\cite{verma2024}. 

\begin{figure}[ht]
    \includegraphics[width=0.8\columnwidth]{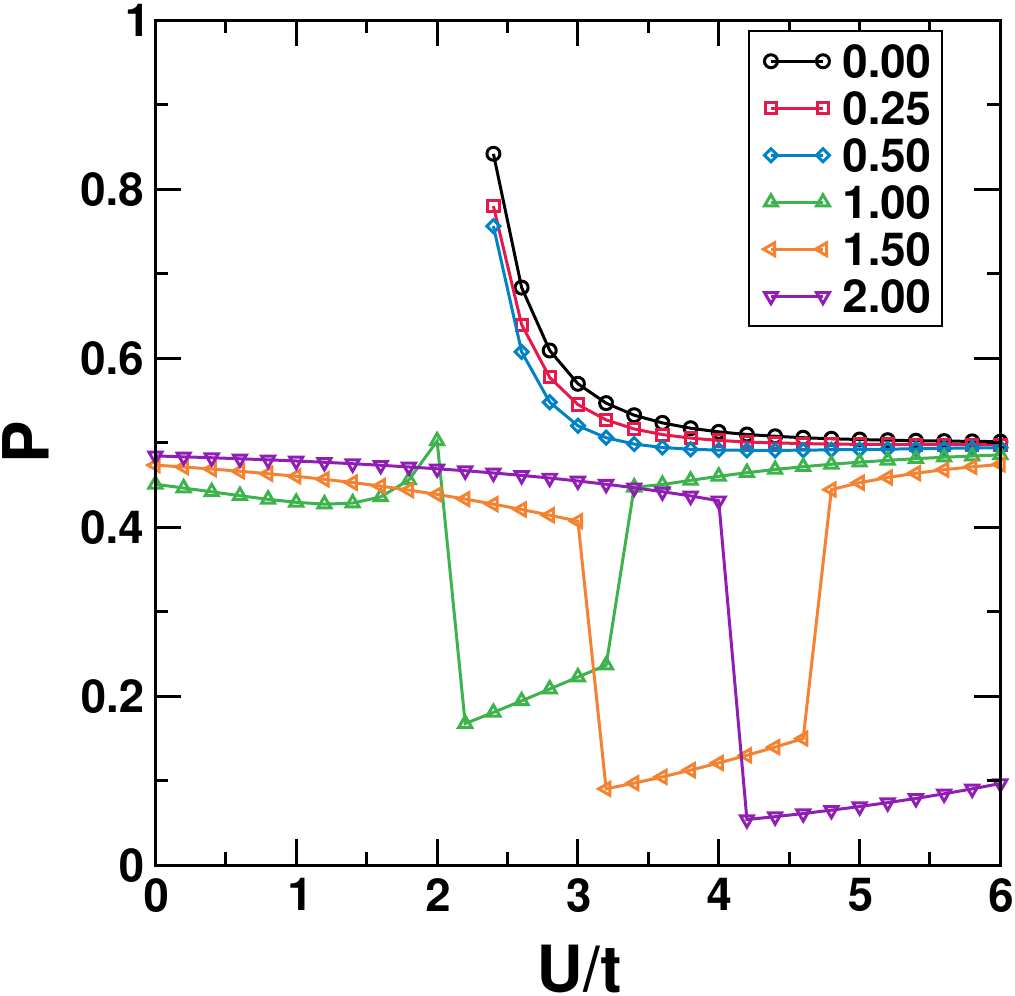}
    \caption{\label{fig:polarization} Variation of Polarization as a function $U$ 
	for $V$ = 0.0, 0.25, 0.5, 1.0, 1.5 and 2.0.}
\end{figure}

\noindent The calculated polarization, $P$ of the system as a function of $U$ 
at different $V$ values is presented in figure \ref{fig:polarization}. 
The system exhibits a metallic ground state for low $U$ and $V$ values. 
Valentim and coworkers have calculated the polarization of the 5/7 skewed ladder 
for $U/t$ values in the range 0 - 5, in the absence of $V$~\cite{Garcia}. 
It should be noted that our band structure calculations 
revealed that the system is metallic in the range 0 $\le$ $U$ $\le$ 2.3 for $V$ = 0.
In the metallic state the polarization of the system is expected to be zero. 
Hence we have not calculated $P$ at these values and have shown only in the 
phases with non-zero energy band gap. 
The $V$ = 0 behavior of the 
$P$ {\it vs}. $U$ plot follows the charge-order phase transition 
as a function of $U$ at a given $V$. For instance, when $V$ = 0, the non-zero polarization 
value appears at $U\approx$ 2.3. The value of $P$ gradually decays to zero 
as we increase $U$. A similar behaviour is observed for $V$ = 0.25 and 0.5 also. 
When $V$ is sufficiently large, say 1.0 or more, we observe multiple transitions in 
$P$ as a function of $U$. All these transitions coincide with the charge-ordering 
transition of the 5/7 skewed ladder. For instance, in case of $V$ = 1.0 (or 2.0), 
we observe transitions at $U \approx$ 2.0 and 3.3 (or 3.0 and 4.7), which correspond to 
the FCO+NM to PCO+PAFM and PCO+PAFM to PCO+SS CFiM phase transitions respectively. 
In case of $V$ = 2.0, we observe a single transition in the polarization curve, 
which corresponds to the FCO+NM to PCO+PAFM phase transition. 
The non-zero polarization in the PAFM or SS CFiM phases indicate 
the presence of multiferroicity in the 5/7 skewed ladder system. 

\section{Discussions}
Conventional antiferromagnetic materials possess net zero magnetization 
and hence are not useful for technological applications. This 
is reflected in their electronic structure as the degeneracy of 
the up- and down-spin bands. Rashba had proposed a momentum dependent 
spin splitting in materials with strong spin-orbit 
interactions ~\cite{pekar}. In contrast, \v Smejkal and 
coworkers through group theoretical studies have shown 
momentum dependent spin splitting in ``altermagnets'' 
possessing certain magnetic symmetry, even in the absence 
of relativistic effects ~\cite{smejkal,smejkal2022}. This spin splitting can enable these 
materials to exhibit interesting properties for technological applications 
like spin current, spin torque, anomalous Hall 
effect etc ~\cite{naka,bai,smejkal2020}. 
It should be noted that both Rashba effect and 
altermagnets show spin splitting in 
a particular momentum direction in the Brillouin zone. Recently, 
SS CFiM with unequal site moment but 
with zero net magnetization has been shown in 
systems like GaFeO\textsubscript{3} ~\cite{dong}. 
Besides, SS AFM 
has also been proposed in MnXN\textsubscript{2} (X = Si, Ge, Sn) 
~\cite{yuan}. Unlike the Rashba effect or altermagnets, 
the SS CFiM and SS AFMs show spin splitting in the entire Brillouin zone. 
Even though the SS CFiM and SS AFM alleviate the 
stringent symmetry requirements of altermagnets, 
their observation in materials is very rare. 
In this context the spin splitting observed in 
our 5/7 skewed ladder in the PCO+SS CFiM phase gains significance. 
It should be noted that in this case the up- and down-spin magnetic sublattices are 
not connected by translation or rotational symmetry unlike the 
conventional antiferromagnets or altermagnets. 

\noindent In a conventional AFM, the combination of 
fractional lattice translational symmetry or spatial inversion symmetry 
and the time-reversal symmetry (TRS) 
is conserved. But, in our 5/7 skewed ladder, in the ground state, the spatial inversion 
symmetry is broken and hence gives rise to spin splitting of the 
energy bands. The absence of the spatial-inversion symmetry also 
gives rise to net non-zero polarization in the lattice in the PCO + SS CFiM phase. 
The coexistence of 
SS and the non-zero polarisation in the PCO + SS CFiM phase makes this a 
SS multiferroic. SS AFMs are technologically 
important as they will enable us to control the 
magnetic property by the application of electric field. 
A similar observation is also made in terms of the 
fully charge-ordered phases FCO1 and FCO2, which 
are degenerate.

\section{Conclusions}
In this work, we employed an extended Hubbard model at half-filling and
explored the rich landscape of electronic and magnetic ground states of
a 5/7 skewed ladder within a mean-field framework. By tuning the on-site
(U) and inter-site (V) Coulomb repulsions, we constructed a phase diagram
that revealed multiple quantum phases in the U-V parameter space.  Using 
the unrestricted mean-field method, we identify various electronic and magnetic phases 
in the U-V plane, by calculating the site charge and spin densities and 
presented the ground state phase diagram of the model. The quantum phase 
diagram consists of multiple competing phases including the partially and 
fully charge-ordered insulators, spin-density wave states, compensated 
ferromagnets with spin-split bands, and non-magnetic metallic regimes. 
Notably, we identified a unique spin-split compensated ferrimagnetic phase 
that exhibits both spin polarisation and finite Berry phase polarisation due 
to the explicit broken inversion symmetry, indicative of a multiferroic behaviour. 
The coexistence of, spin polarisation and net electronic polarisation in
this phase highlights its potential relevance to spintronics and multifunctional 
device applications. Our work thus offers critical insight into the interplay 
between charge, spin, and symmetry in low-dimensional systems, particularly 
those modeling realistic grain boundary geometries like the \{112\}  boundary in silicon.

\begin{acknowledgments}
SJ and RR acknowledge Dr. Ram Janay Choudhary, UGC-DAE CSR Indore for some insightful discussions. 
\end{acknowledgments}

\vspace{1em}
\centerline {\bf Data Availability}
The data that support the findings of this article are not
publicly available. The data are available from the authors upon
reasonable request.
%=======================================================
%=======================================================
%\twocolumngrid
\bibliographystyle{apsrev4-2}  % Use the Physical Review style for references
\bibliography{references}  % Load your .bib file here
\end{document}